\documentclass[prl,aps,twocolumn]{revtex4}
\usepackage{graphicx}
\usepackage{amsmath}
\usepackage{bm}
\usepackage{amstext}
\usepackage{amsxtra}
\usepackage{gensymb}

\usepackage{times}

\begin{document}

\newcommand{\To}{T_c^0}
\newcommand{\kB}{k_{\rm B}}
\newcommand{\dT}{\Delta T_c}
\newcommand{\lo}{\lambda_0}
\newcommand{\cs}{$\clubsuit$}
\newcommand{\thold}{t_{\rm hold}}
\newcommand{\Nmf}{N_c^{\rm MF}}
\newcommand{\Neq}{N_0^{\rm eq}}
\newcommand{\Tmf}{T_c^{\rm MF}}
\newcommand{\el}{\gamma_{\rm el}}
\newcommand{\meq}{\mu^{\rm eq}}
\newcommand{\K}{^{39}{\rm K}}
\newcommand{\Teq}{T^{\rm eq}}

\title{A superheated Bose-condensed gas}

\author{Alexander L. Gaunt*, Richard J. Fletcher*, Robert P. Smith$^\dag$, and Zoran Hadzibabic}
\affiliation{Cavendish Laboratory, University of Cambridge, J.~J.~Thomson Ave., Cambridge CB3~0HE, United Kingdom}

\begin{abstract}
%We have observed and explored a superheated atomic Bose gas, in which a Bose-Einstein condensate (BEC)  persists above the equilibrium critical temperature if it is decoupled from the decohering thermal bath by reducing the strength of interatomic interactions.
%For vanishing interactions the BEC persists in the superheated regime for a minute, and for a total number of atoms in the gas three times smaller than the equilibrium critical value.
%If strong interactions are suddenly turned on, the BEC rapidly ``boils" away.
%We reconstruct a non-equilibrium phase diagram of this system and reproduce our observations in numerical simulations.
\end{abstract}

\date{\today}


%03.75.Hh	Static properties of condensates; thermodynamical, statistical, and structural properties
%67.85.-d 	Ultracold gases, trapped gases
%03.75.Kk 	Dynamic properties of condensates; collective and hydrodynamic excitations, superfluid flow
%Optical angular momentum (quantum optics), 42.50.Tx
%47.37.+q 	Hydrodynamic aspects of superfluidity; quantum fluids
%67.85.De 	Dynamic properties of condensates; excitations, and superfluid flow
%37.10.Vz 	Mechanical effects of light on atoms, molecules, and ions

\maketitle

{\bf
Our understanding of various states of matter usually relies on the assumption of thermodynamic equilibrium. 
However, the transitions between different phases of matter can be strongly affected by non-equilibrium phenomena.
Here we demonstrate and explain an example of non-equilibrium stalling of a continuous, second-order phase transition. We create a superheated atomic Bose gas, in which a Bose-Einstein condensate (BEC)  persists above the equilibrium critical temperature, $T_c$, if its coupling to the surrounding 
thermal bath is reduced by tuning interatomic interactions.
For vanishing interactions the BEC persists in the superheated regime for a minute.
However, if strong interactions are suddenly turned on, it rapidly ``boils" away.
Our observations can be understood within a two-fluid picture, treating the condensed and thermal components of the gas as separate equilibrium systems with a tuneable inter-component coupling. 
We experimentally reconstruct a non-equilibrium phase diagram of our gas, and theoretically reproduce its main features.
}

Non-equilibrium many-body states can persist for a very long time if,
for example, a system is integrable, the transition to the lower free-energy state is inhibited by an energy barrier, or the target equilibrium state is continuously evolving due to dissipation. Ultracold atomic gases offer excellent possibilities for fundamental studies of non-equilibrium phenomena \cite{Polkovnikov:2011,Kinoshita:2006,Winkler:2006,Sadler:2006,Hofferberth:2007,Haller:2009,Guzman:2011,Cheneau:2012,Trotzky:2012,Mark:2012,Smith:2012,Gring:2012} and have been used to create counter-intuitive states such as repulsively bound atom pairs \cite{Winkler:2006} and Mott insulators with attractive inter-particle interactions \cite{Mark:2012}.

\begin{figure*} [btp]
\includegraphics[width=\textwidth]{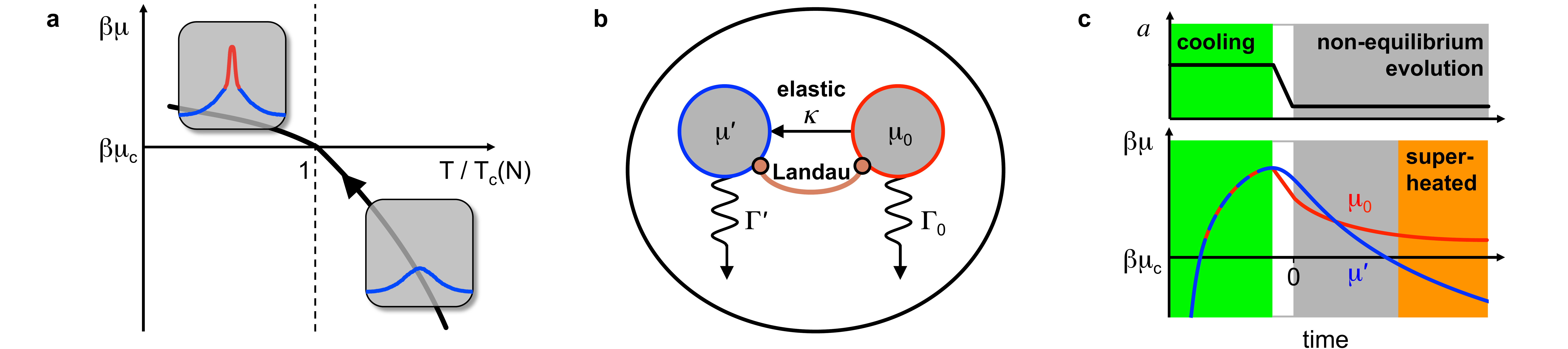}
\caption{Creating and understanding a superheated Bose-condensed gas. (a) In equilibrium, a BEC is present if $T<T_c$ or equivalently $\mu > \mu_c$ (here $\beta = 1/(\kB T)$). The arrow indicates the cooling trajectory along which a BEC is produced. The insets show measured momentum distributions, with the condensed component indicated in red. 
(b) Two-component picture. The thermal and condensed components have chemical potentials $\mu'$ and $\mu_0$, and inelastic decay rates $\Gamma'$ and $\Gamma_0$, respectively. The net flow of particles between the two components, $\kappa$, depends on $\mu'$, $\mu_0$ and the scattering rate $\propto a^2$. In equilibrium $\mu' = \mu_0 = \meq$. The Landau damping of the collective modes in the BEC has a rate $\propto \sqrt{a}$.
(c) Time-sequence of the experiment. Reducing $a$ after preparing a BEC reduces the coupling between the two components and extends the condensate lifetime. In the superheated regime $\mu' < \meq < \mu_c$ but $\mu_0 > \mu_c$. 
}
\label{fig:basicidea}
\end{figure*}

Our superheated Bose gas is reminiscent of superheated distilled water, which remains liquid above 100 $\degree$C. 
Specifically, as the temperature characterising the average energy per particle and the populations of the excited states rises above $T_c$, the cloud remains in the partially condensed phase, which in true equilibrium should exist only below $T_c$.
However, there are also important differences. 
Boiling of water is a first-order phase transition and is stalled in clean samples by the absence of nucleation centres.
In that case the transition is inhibited by an energy barrier.
For a second-order phase transition such a barrier does not exist and the superheating we observe is a purely dynamical non-equilibrium effect, which arises because different properties of the system evolve at different rates. 
In this respect our gas also bears resemblance to the long-lived non-equilibrium spin structures observed in spinor condensates \cite{Sadler:2006,Guzman:2011}, pre-thermalised states in quenched one-dimensional (1D) Bose gases \cite{Gring:2012} and supercritical superfluids predicted to occur in quenched 2D gases \cite{Mathey:2010}. In all those cases, however, non-equilibrium states are observed due to the system's {\it slow approach} to a true equilibrium. 
Here, the system actually evolves {\it away} from equilibrium.

In Fig.~\ref{fig:basicidea} we summarise the basic idea of our experiments and the key concepts needed to understand them. In an equilibrium gas, a BEC is present only if $T<T_c$, where $T_c$ depends on the total particle number $N$, or equivalently if the chemical potential is $\mu > \mu_c$. 
In a standard experiment, after a BEC is produced, it gradually decays because $T$ rises, due to technical heating, and/or $T_c$ decreases, because $N$ decays through various inelastic processes. 
As $T/T_c$ increases, elastic collisions redistribute the atoms between the thermal and condensed components, aiming to ensure the equilibrium particle distribution. 
The BEC atom number, $N_0$, can therefore decay in two ways: (1) by direct inelastic loss, and (2) through elastic transfer of atoms into the thermal component. Here we reduce the rate of the elastic particle transfer by tuning the strength of inter-particle interactions, characterised by the $s$-wave scattering length $a$. 
This protects the BEC deep into the superheated regime, where $N_0 > 0$ even though $T>T_c$. 

We can understand our observations within the two-fluid picture outlined in Fig.~\ref{fig:basicidea}(b).
Here we treat the thermal and condensed components as two coupled sub-systems with atom numbers $N'$ and $N_0$, chemical potentials $\mu' $ and $\mu_0$, and instantaneous per-particle inelastic decay rates $\Gamma' $ and $\Gamma_0$, respectively. In equilibrium $\mu' = \mu_0$; note that $\mu_0$ is defined only if $N_0>0$, so $\mu_0 > \mu_c$.

The two components are coupled in two ways, both dependent on the scattering length $a$. First, the local ``kinetic" thermal equilibrium between the collective excitations in the BEC (phonons) and the thermal bath is ensured by Landau damping, the rate of which is $\propto \sqrt{a}$ \cite{Fedichev:1998,Pethick:2002,SuperheatedSI}. Second, the global ``phase" equilibrium (i.e., the equilibrium condensed fraction $N_0/N$) is ensured by the elastic scattering with a rate $\propto a^2$.
Crucially, due to the different scalings with $a$, we find a large parameter space where the two components can be considered to be in local kinetic equilibrium while the system is not in global phase equilibrium. In other words, the two components are at the same temperature, but have different chemical potentials. 

In our optically trapped $\K$ gas \cite{Campbell:2010}, we control $a$ by an external magnetic field tuned close to a Feshbach resonance at 402 G \cite{Roati:2007}, the dominant source of $\Gamma'$ and $\Gamma_0$ is spontaneous scattering of photons from the trapping laser beams, and $\Gamma_0$ has an additional contribution from three-body recombination.

The key steps in our experimental sequence are summarised in Fig.~\ref{fig:basicidea}(c). We start by preparing a partially condensed gas in the $|F,m_F\rangle = |1,1\rangle$ hyperfine ground state by evaporative cooling at $a=135\,a_0$, where $a_0$ is the Bohr radius \cite{Campbell:2010}. 
We then reduce $a$ (over 50 ms) 
and follow the subsequent evolution of the cloud, probing the atomic momentum distribution by absorption imaging in time-of-flight expansion.
Reducing $a$ (at constant $N_0$) initially reduces $\mu_0$ below $\mu'$ \cite{Smith:2012}, but subsequently $\mu_0$ decays slower.

\begin{figure} [bp]
\includegraphics[width=\columnwidth]{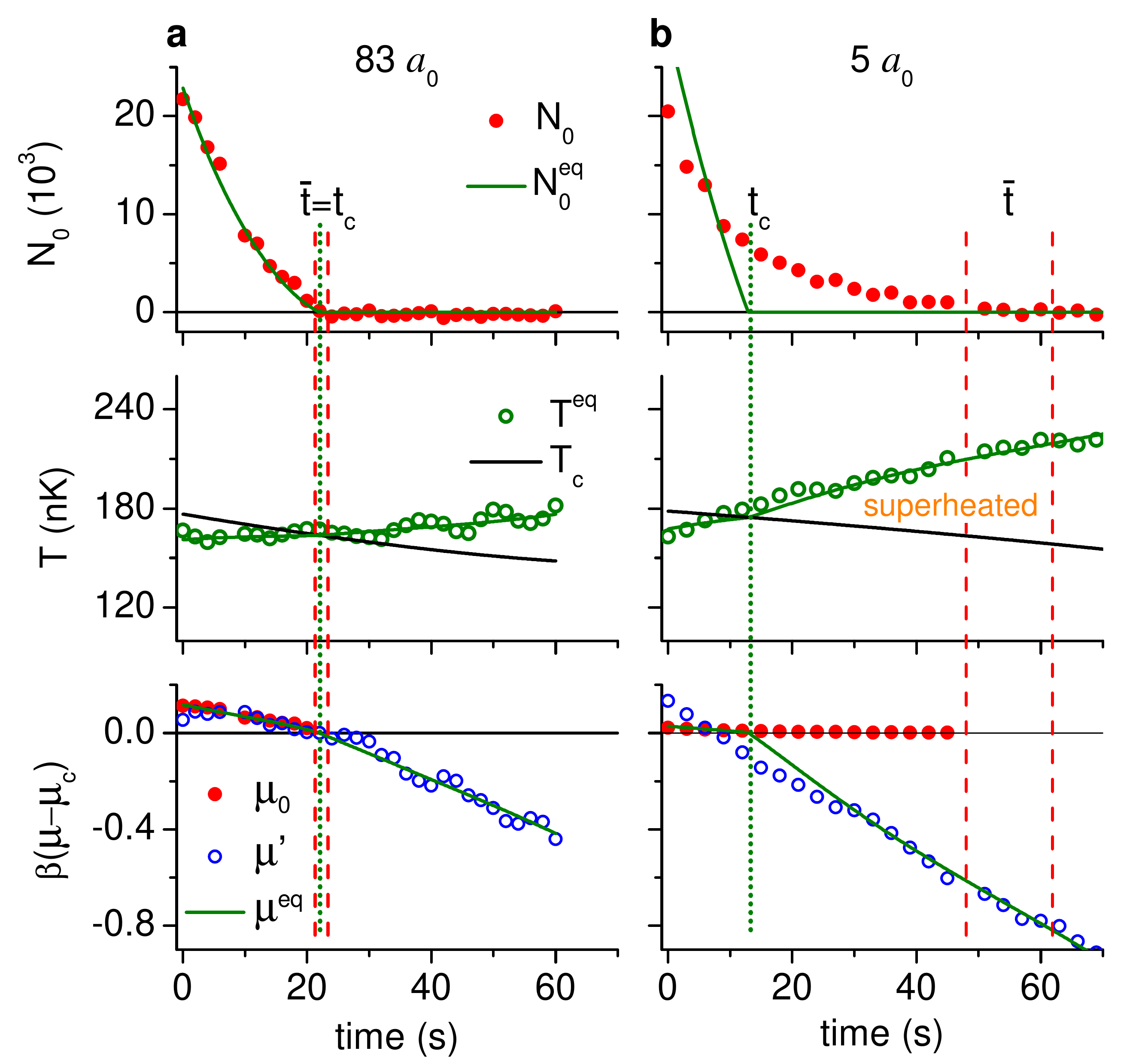}
\caption{Equilibrium vs. non-equilibrium BEC decay. (a) At $a=83\,a_0$ the cloud is always in quasi-static equilibrium. The measured $N_0$ is in excellent agreement with the predicted $\Neq$ and vanishes when $\Teq = T_c$; the three separately calculated chemical potentials, $\mu_0$, $\mu'$ and $\meq$, all agree with each other. The dotted green line marks the equilibrium critical time, $t_c$, and the dashed red lines show the experimental bounds on the time $\bar{t}$ when the BEC actually vanishes. 
(b) At $5\,a_0$, the BEC persists in the superheated regime ($\Teq>T_c$) for $\bar{t} - t_c \approx 40\,$s. 
}
\label{fig:process}
\end{figure}

In Fig.~\ref{fig:process} we quantitatively contrast the equilibrium evolution of a cloud at $a=83\,a_0$ and the non-equilibrium evolution at $5\,a_0$. In both cases we start at time $t=0$ [see Fig.~\ref{fig:basicidea}(c)] with $N_0 \approx 2 \times 10^4$ and $N \approx 2 \times 10^5$ at $T \approx 160\,$nK.
In both cases $T_c$ decreases at a similar rate due to similar $N$-decay. At $5\,a_0$, the temperature rises faster due to less effective evaporative cooling at a fixed optical trap depth.

Whether the gas is in equilibrium or not, it can always be characterised by two {\it extensive} variables, the total particle number $N$ and energy $E$. We measure these quantities by direct summation of the momentum distribution and its second moment. To measure $N_0$ we count the atoms within the central peak rising above the smooth thermal distribution.

From the measured $N(t)$ alone we calculate the equilibrium $T_c(t)$ \cite{Smith:2011}.
From $N(t)$ and $E(t)$ we calculate the equilibrium {\it intensive} thermodynamic variables $\meq (t)$ and $\Teq(t)$, and the equilibrium number of condensed atoms, $\Neq (t)$ \cite{Smith:2011,Tammuz:2011, Smith:2012b}; in these calculations $\Neq>0$ if and only if $\Teq < T_c$. For comparison, we also directly fit a temperature $T^f$ to the wings of the momentum distribution.
Additionally, supposing only that the two components are separately in equilibrium, from the measured $N_0$ and $N'$ we calculate $\mu_0$ and $\mu'$. 
(For theoretical details see Supplementary Information \cite{SuperheatedSI}.)

At $83\,a_0$ [Fig.~\ref{fig:process}(a)] we find excellent agreement between the measured $N_0$ and the $\Neq$ predicted without any free parameters. The BEC vanishes exactly at the equilibrium ``critical time" $t_c$ (dotted green line), at which $\Teq = T_c$. Note that the dashed red lines show the experimental bounds on the time $\bar{t}$ when the BEC vanishes.
The separately calculated $\mu_0$, $\mu'$ and $\meq$ are all consistent and we have also checked that the fitted $T^f$ coincides with the calculated $\Teq$. All this gives us full confidence in our equilibrium calculations.

At $5\,a_0$ [Fig.~\ref{fig:process}(b)] we observe strikingly different behaviour. The BEC now survives much longer than it would in true equilibrium; $\bar{t} - t_c \approx 40\,$s. 
We also see that $\mu_0$ and $\mu'$ diverge from each other for $t>t_c$, so the system is moving {\it away} from the global phase equilibrium rather than {\it towards} it. The observed superheating can thus not be understood as just a transient effect. 
(Note that $\mu_0 - \mu_c$ is always very small due to weak interactions.)

At $5\,a_0$ the gas is not in global phase equilibrium, 
but it is still a good approximation to view its two components as two equilibrium sub-systems at a same (kinetic) temperature, as in Fig.~\ref{fig:basicidea}(b).
We have checked that the momentum distribution in the non-condensed component is still fitted well by a thermal distribution, with $T^f$ always within $10\,\%$ of the calculated $\Teq (N,E)$
(see Supplementary Information \cite{SuperheatedSI}).
For the BEC, in a weakly interacting gas the equilibrium relation $\mu_0 (N_0)$ relies on the macroscopic occupation of a single quantum state \cite{Dalfovo:1999}, rather than on global equilibrium. 
Moreover, even for the lowest-energy collective modes we estimate the Landau damping time to be $<1\;$s~\cite{Fedichev:1998,Pethick:2002,SuperheatedSI}, i.e. much shorter than the characteristic time scale of our experiments. Thus, while this distribution is not directly measurable, we expect the distribution of collective excitations in the BEC to be characterised by a temperature $T_0$ that is also close to $\Teq \approx T^f$.

These conclusions hold for any $a \gtrsim 1\,a_0$~\cite{SuperheatedSI}. Exactly at $a=0$ our theoretical picture does break down, since the Landau damping rate vanishes and the BEC has no equilibrium features; the two components are simply completely decoupled. Bearing this small caveat in mind, from here on we refer to $T_0 \approx T^f \approx \Teq$ simply as  the temperature of the system $T$.

\begin{figure} [btp]
\includegraphics[width=\columnwidth]{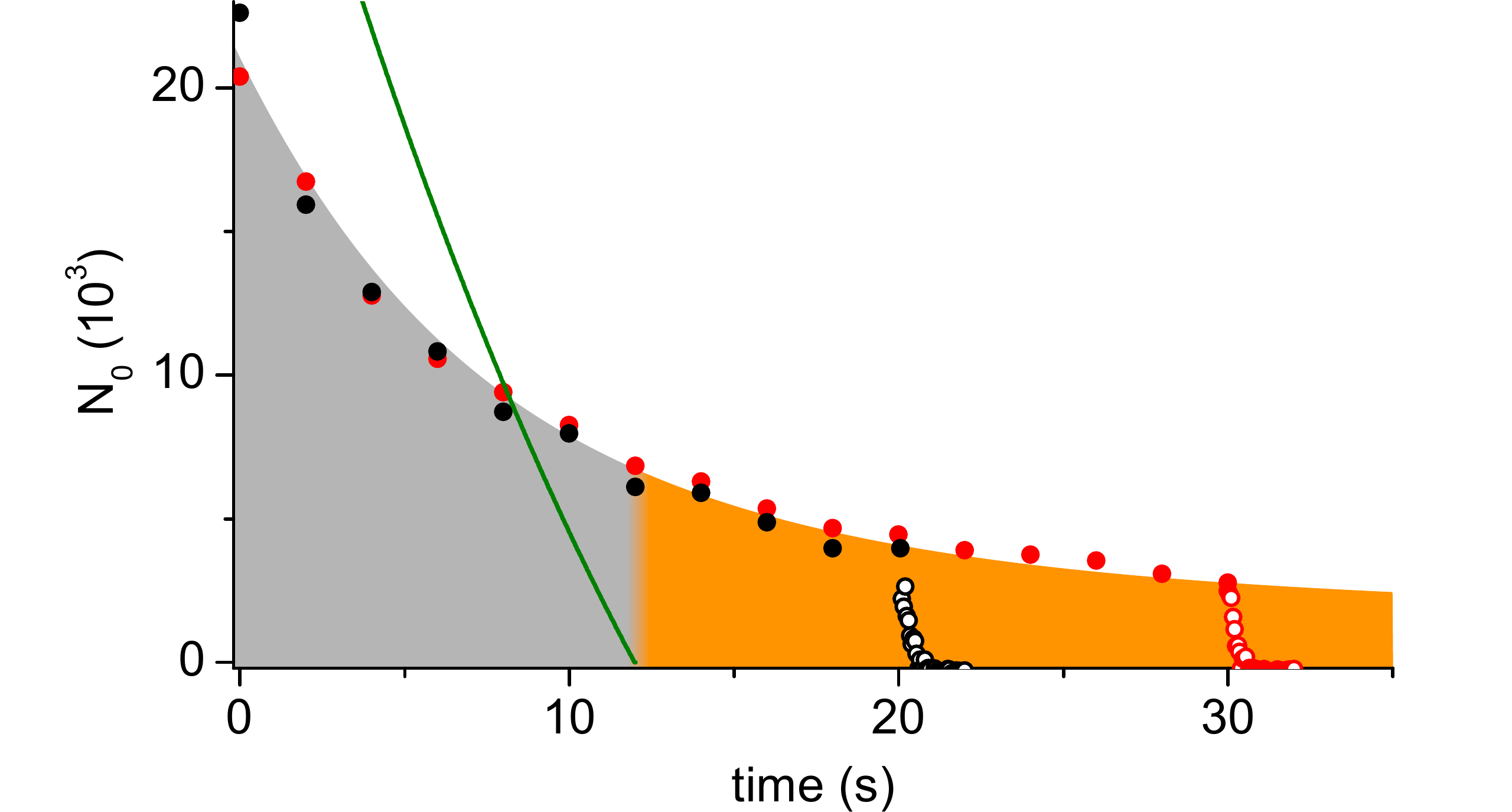}
\caption{Quenching the superheated Bose-condensed gas. Solid symbols show the evolution of $N_0$ at $a=3\,a_0$, 
the green solid line shows $\Neq$ and orange shading indicates the superheated regime. 
Open symbols show the rapid decay of the BEC after it is strongly coupled to the thermal bath by an interaction quench to $a=62\,a_0$ at time $t_q$. We show two experimental series in which $t_q = 20\,$s (black) and 30~s (red). 
}
\label{fig:highway}
\end{figure}

If superheated water is perturbed, e.g., by sprinkling some salt into it, it rapidly boils away.
Here, an analogous way to directly see that the gas is superheated is to suddenly increase the coupling of the BEC to the thermal bath. In Fig.~\ref{fig:highway} we show the results of two experimental series in which $a$ is quenched (within 10~ms) from $3\,a_0$ to $62\,a_0$ at different times in the superheated regime. The solid (open) symbols show $N_0$ measured before (after) the quench. The small $\Gamma_0$ is essentially unaffected by the change in $a$, and the sudden $N_0$ decay is due to the increase in $\kappa$ [see Fig.~\ref{fig:basicidea}(b)].  For reference, the green line shows the calculated $\Neq$ at $3\,a_0$ and orange shading indicates the superheated regime.

\begin{figure} [tbp]
\includegraphics[width=\columnwidth]{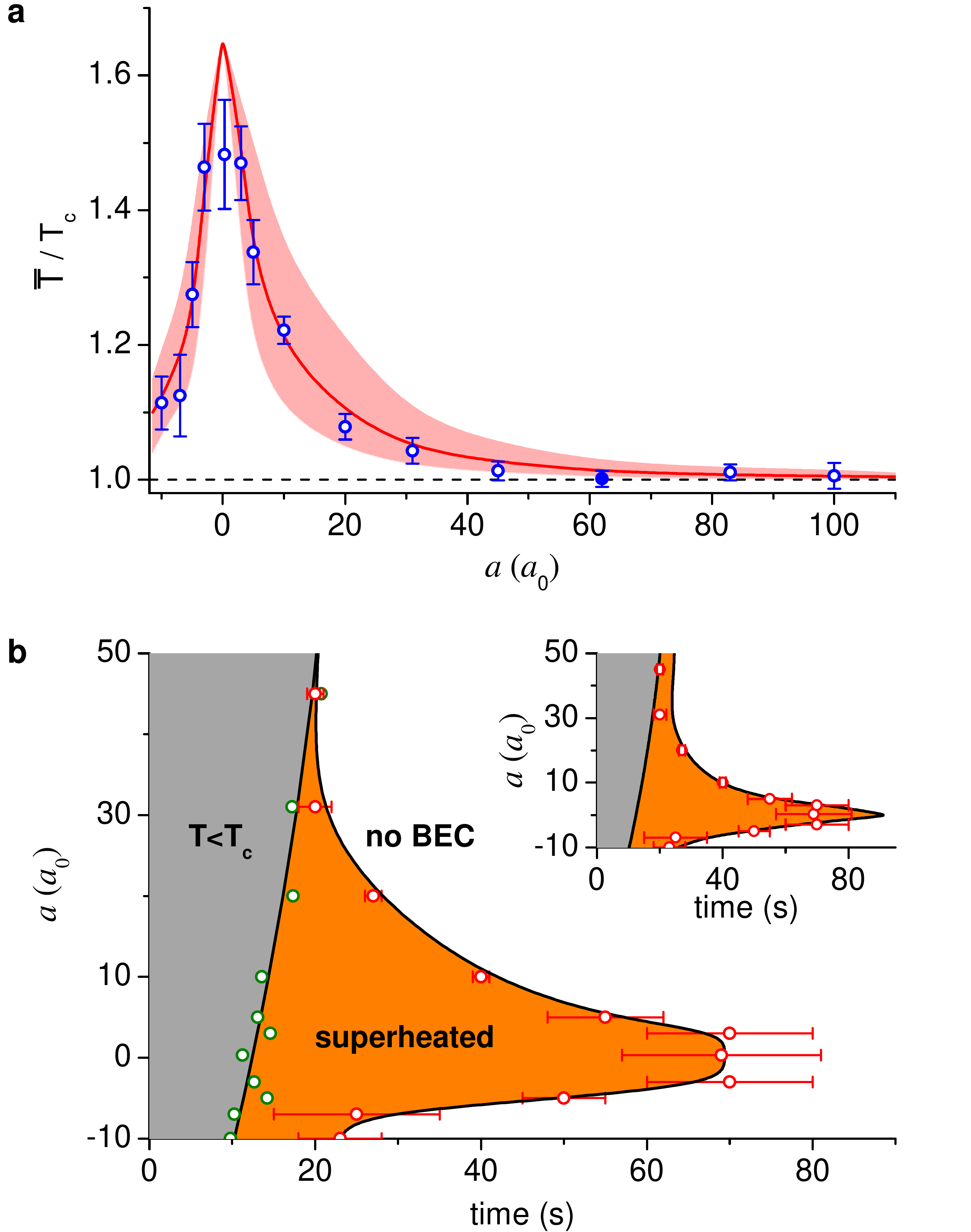}
\caption{Limits of superheating. (a) The highest temperature at which we observe a BEC, $\bar{T}$, scaled to the equilibrium $T_c$. Close to $a=0$ the BEC survives up to $\approx 1.47 \,T_c$. The red line shows the results of our numerical calculations, with the shaded area indicating the theoretical uncertainty. Experimental error bars are statistical. The point at $62\,a_0$ is fixed to unity by the absolute atom number calibration \cite{Smith:2011, Smith:2011b}.
(b) Temporal phase diagram. For each value of $a$ we plot the equilibrium $t_c$ (green points) and the time $\bar{t}$ at which the BEC actually vanishes (red points). The $\bar{t}$ errors correspond to dashed lines in Fig.~\ref{fig:process} and the uncertainty in $t_c$ is indicated by the scatter of points. Solid curves are spline fits to the data. For $a\approx 0$ the BEC survives in the superheated regime for a whole minute. Inset: numerically calculated phase diagram, with $\bar{t}$ data overlaid.
}
\label{fig:master}
\end{figure}

As shown in Fig.~\ref{fig:master}, we have explored the limits of superheating for a range of interaction strengths, including small negative values of $a$. For $a <0$, a BEC is stable against collapse only for $N_0 < -C/a$, with $C \approx 2 \times 10^4 \, a_0$ for our trap parameters \cite{Ruprecht:1995, Gerton:2000,Donley:2001}. However, after $N_0$ drops below this critical value, at small $|a|$ it decays slowly.

In Fig.~\ref{fig:master}(a) we plot the highest temperature at which we still observe a BEC, $\bar{T} \equiv T(t=\bar{t})$, scaled to the equilibrium $T_c$ at the same $N$. For $a \rightarrow 0$, the BEC survives up to $T \approx 1.47\,T_c$. 
(For comparison, this is analogous to superheated water at $275\,\degree$C.)

In Fig.~\ref{fig:master}(b) we reconstruct the temporal phase diagram of our non-equilibrium gas. Here, a horizontal cut through the graph corresponds to a time series such as shown in Fig.~\ref{fig:process}.
For each $a$, we plot the measured $\bar{t}$ (red points) and the equilibrium $t_c$ (green points). The solid curves are spline fits to the data. The width of the orange-shaded region corresponds to the time that the BEC survives in the superheated regime. For $a \approx 0$ this region spans a whole minute.

The phase diagram in Fig.~\ref{fig:master}(b) is measured by always starting with $N_0 \approx 2 \times 10^4$. In general, non-equilibrium behaviour can strongly depend on the initial conditions. However, we find that $\bar{t}$ is essentially constant (within experimental errors) for initial $N_0$ in the range $(1-5)\times 10^4$. The primary reason for this is that the three-body contribution to $\Gamma_0$ grows with $N_0$; this leads to ``self-stabilisation" of the condensed atom number on timescales much shorter than $\bar{t}$.

We theoretically reproduce our non-equilibrium observations using
a two-component model directly corresponding to Fig.~\ref{fig:basicidea}(b). Starting with the measured initial $N_0$, we numerically simulate the evolution of a BEC coupled to a thermal bath characterised by $\mu'(t)$. To do this we calculate $\Gamma_0$ from our experimental parameters,  and for $\kappa$ we use the form \cite{Gardiner:1997a}
\begin{equation}
\kappa = A \el N_0 \left[e^{\beta (\mu_0 - \mu_c)} - e^{\beta (\mu' - \mu_c)} \right] \, .
\label{eq:kappa}
\end{equation}
Here $\el \propto a^2$ is the elastic collision rate
and $A$ is a dimensionless coefficient. The largest uncertainty in our calculations comes from the theoretical uncertainty in $A \approx 1-10$ \cite{Gardiner:1998a}. (For details see Supplementary Information \cite{SuperheatedSI}.)

In Fig.~\ref{fig:master}(a) we show the calculated $\bar{T}/T_c$. The red line corresponds to $A=3$ and the shaded area to the range $A=1-10$.
The calculation generally captures our experimental observations well. With $A=3$ we obtain quantitative agreement with the data, except exactly at $a=0$, where the model is not valid.
In the inset of Fig.~\ref{fig:master}(b) we show the calculated temporal phase diagram, with $A=3$, together with the experimental $\bar{t}$ data. Again the general features of the diagram are captured well for $a \neq 0$.

In conclusion, we have observed superheating in a Bose-condensed gas with tuneable interactions, mapped out a non-equilibrium phase diagram of this system, and reproduced our measurements in numerical simulations based on a two-fluid picture of a partially condensed gas.
The success of our calculations supports a conceptually simple way to think about dynamical non-equilibrium effects near  a continuous phase transition. Extending the BEC lifetime by tuning interactions could also have practical benefits for precision measurements and quantum information processing.

%%%%%%%%%%%%%%%%%%%%%

%\acknowledgments{
%We thank S. Beattie and S. Moulder for experimental assistance and critical reading of the manuscript. 
%This work was supported by EPSRC (Grant No. EP/I010580/1), AFOSR, ARO and DARPA OLE.}

%*rps24@cam.ac.uk

%\bibliography{Quench}
%\bibliographystyle{Nature}

%%%%%%%%%%%%%%%%%
\vspace{5mm}

\noindent * These two authors contributed equally.

\noindent $^\dag$ rps24@cam.ac.uk

\noindent \textbf{Acknowledgements} 

\noindent We thank S. Beattie and S. Moulder for experimental assistance. 
This work was supported by EPSRC (Grant No. EP/K003615/1), the Royal Society, AFOSR, ARO and DARPA OLE.

\noindent \textbf{Author Contributions} 

\noindent All authors contributed extensively to this work.

\noindent \textbf{Additional Information} 

\noindent The authors declare no competing interests.
Correspondence and requests for materials should be addressed to R. P. S.

%%%%%%%%%%%%%%%%%%%%%%%%%%%%%%%%%%%%%
%%%%%%%%%%%%%%%%%%%%%%%%%%%%%%%%%%%%%%

\onecolumngrid
\newpage

\begin{center}
\textbf{Supplementary Information}
\end{center}

\vspace{2mm}

\renewcommand{\figurename}{Fig. S\!\!}

\subsection{Determination of $N$, $E$ and $N_0$}

We take an absorption image of the atom cloud after $\tau= 18$~ms of time-of-flight (TOF) expansion from a nearly isotropic trap with a geometric mean of trapping frequencies $\bar{\omega}/2\pi \approx 70\,$Hz.

For absolute calibration of our atom numbers we use a $T_c$ measurement at $a=62\,a_0$, assuming that at this $a$ the cloud is in equilibrium \cite{Smith:2011SI, Smith:2011bSI}.

For $a<100\,a_0$ we assess the interaction-energy contribution to the total energy $E$ to be $\lesssim 1\%$ and thus $E \approx 2 E_k$, where $E_k$ is the kinetic energy. We obtain $E_k$ from the second moment of the atom distribution after TOF, and correct it for the small effect of the initial in-trap cloud size. This amounts to rescaling the energy by a factor $\bar{\omega}^2\tau^2/(1+\bar{\omega}^2\tau^2)$.

\begin{figure} [h]
\includegraphics[width=0.5\columnwidth]{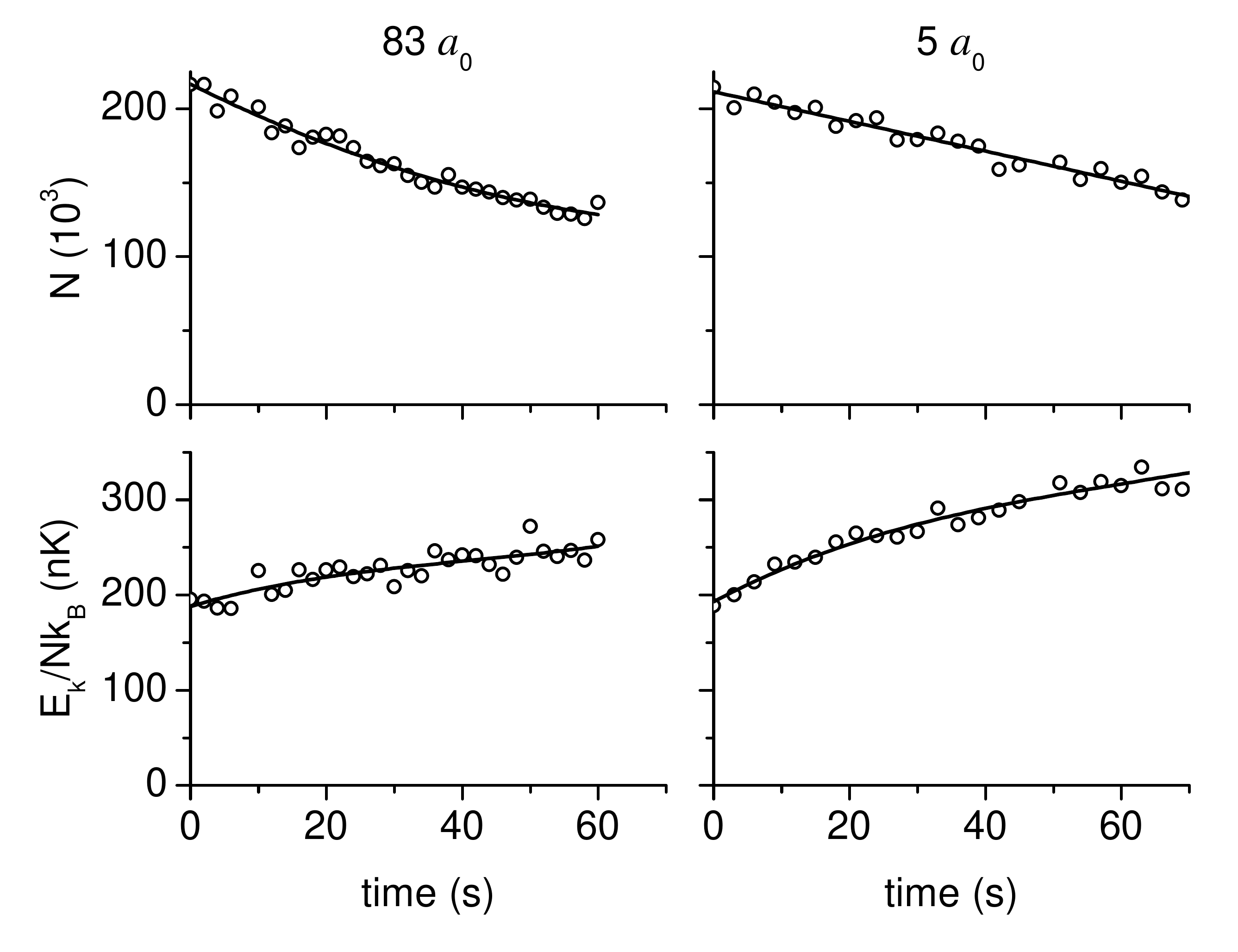}
\caption{\textbf{Parametrization of $N$ and $E_k$}. The total atom number $N$ and kinetic energy per particle $E_k/N$ for two experimental series at $83\,a_0$ and $5\,a_0$.  The experimental data are fitted with polynomial forms to obtain smooth functions $N(t)$ and $E_k(t)$.}
\label{fig:NandE}
\end{figure}

In Fig.~S\ref{fig:NandE} we show $N$ and $E_k$ for the same two experimental series shown in Fig.~2 in the main text.

To improve the detection of small $N_0$ values we always switch $a$ to zero at the start of TOF \cite{Smith:2011SI}.

\subsection{Equilibrium calculations}

For an equilibrium ideal Bose gas in a spherically symmetric harmonic trap of frequency $\omega$, the thermal component satisfies:
\begin{equation}
N'(\mu',T)=\frac{N_c^0}{\zeta(3)}
\int_0^{\infty} \mathrm{g}_2\left(\exp\left(\frac{\mu'-\mu_c^0}{\kB T}-\frac{x^2}{2}\right)\right) x \mathrm{d}x,
\label{eq:N'}
\end{equation}
\begin{equation}
\frac{E_k(\mu',T)}{N'(\mu',T)}=\frac{3}{2} \kB T \frac{
\int_0^{\infty}\mathrm{g}_2\left(\exp(\frac{\mu'-\mu_c^0}{\kB T}-\frac{x^2}{2})\right)\frac{x^2}{2}x \mathrm{d}x}
{\int_0^{\infty}\mathrm{g}_2\left(\exp(\frac{\mu'-\mu_c^0}{\kB T}-\frac{x^2}{2})\right)x \mathrm{d}x},
\label{eq:E}
\end{equation}
where $N_c^0=\zeta(3)\left(\frac{\kB T}{\hbar \omega}\right)^3 \left(1-\frac{\zeta(2)}{2\zeta(3)}\frac{\hbar \omega}{\kB T}\right)^{-3}$ includes the finite size correction, $\mu_c^0=\frac{3}{2}\hbar \omega$, and $\mathrm{g}_2(z)$ is the dilogarithm function.
For $N>N_c^0$, the chemical potential is capped at $\mu'=\mu_c^0$ and also $\mu_0 = \mu_c^0$. The thermal atom number is saturated, $N'=N_c^0$, and any additional particles must go into the condensate, $N_0=N-N_c^0$.

Interactions modify this picture in two ways:

(i) The critical point is shifted. At mean-field (MF) level $\mu_c=\mu_c^0+4\zeta(3/2)a/\lambda$, where $\lambda=\sqrt{\frac{2 \pi \hbar^2}{m \kB T}}$ is the thermal wavelength. A small beyond-MF correction is quadratic in $a/\lambda$ and has an additional logarithmic correction. Experimentally, the corresponding $N_c$ is \cite{Smith:2011SI, Smith:2011bSI}
\begin{equation}
N_c=N_c^0\left(1-3.426\frac{a}{\lambda}+42\left(\frac{a}{\lambda}\right)^2\right)^{-3} \; .
\label{eq:Nc}
\end{equation}

(ii) Due to interactions, in the presence of a BEC, $N'$ is no longer saturated at $N_c$. Empirically,
\begin{equation}
N'=N_c +S_0 (N_0)^{2/5}+S_2 (N_0)^{4/5},
\label{eq:eos}
\end{equation}
where the non-saturation parameters $S_0$ and $S_2$ depend on $a$ and $T$ \cite{Tammuz:2011SI, Smith:2012bSI}. The excess number of thermal atoms, $N' - N_c$, can be directly attributed to the shift of the chemical potential above $\mu_c$; for an interacting BEC $\mu_0 > \mu_c$ and in equilibrium $\mu' = \mu_0$.

For $\mu_0$ we use a modified Thomas-Fermi law \cite{Gardiner:1998aSI}:
\begin{equation}
\mu_0-\mu_c =\frac{\hbar \omega}{2} \left\{\left(15\frac{N_0a}{a_{\rm osc}} + 3^{5/2}\right)^{2/5}-3\right\},
\label{eq:mu0}
\end{equation}
where $a_{\rm osc}=\sqrt{\hbar/m \omega}$ is the harmonic oscillator length.

Eq.~(\ref{eq:mu0}) and Eqs.~(\ref{eq:N'}) and (\ref{eq:E}) modified to include interaction effects ($N_c^0 \rightarrow N_c$ and $\mu_c^0 \rightarrow \mu_c$) form a complete set needed for our calculations \cite{g2SI}.
We proceed in two ways:

(1) Assuming that the system is in global equilibrium, $\mu'=\mu_0=\mu^{\rm eq}$, we use {\it only} $N(t)$ and $E(t)$ to calculate $\mu^{\rm eq}(t)$, $N_0^{\rm eq}(t)$ (green lines in main-text Fig.~2) and $T^{\rm eq}(t)$.

(2) Additionally, from $N$, $E$ and the {\it measured} $N_0$ we calculate $\mu_0$ and $\mu'$ without assuming $\mu'=\mu_0$ (red and blue points in the bottom panel of main-text Fig.~2).

\subsection{Justification of the two-fluid picture for a gas out of global equilibrium}

In our theoretical picture (Fig.~1(b) in the paper), we assume that in the superheated regime the thermal and condensed components can still to a good approximation be assumed to be separately in equilibrium. Moreover, we assume that they are at the same temperature, but just have different chemical potentials.

Here we provide a more detailed justification of these assumptions.

\begin{figure} [h]
\includegraphics[width=\columnwidth]{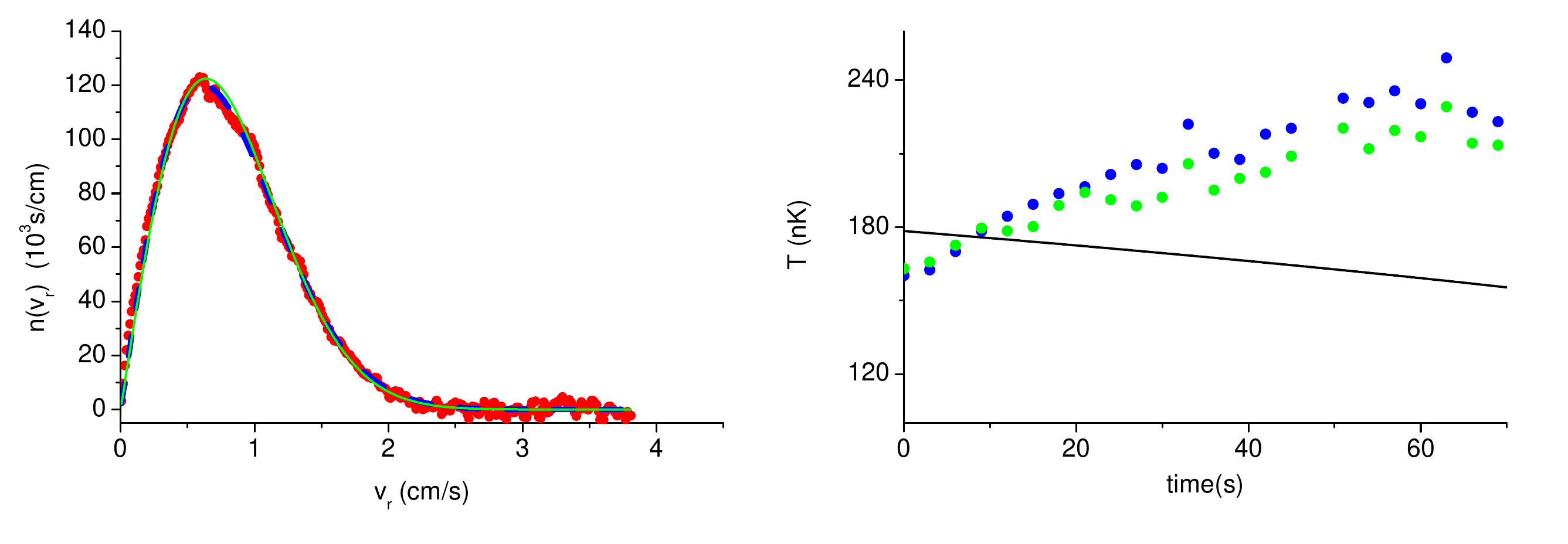}
\caption{\textbf{Thermal distribution in a gas out of global phase equilibrium}. (a) For a gas at $5\;a_0$ in the superheated regime we show the radial velocity distribution (red), the distribution corresponding to $N$, $E$ and $\Teq$ (green) and the unconstrained fit (blue) giving $T^f$. Even though the gas is not in true global equilibrium the distribution still looks thermal and $T^f$ and $\Teq$ agree to within few \%. (b) Comparison of $T^f$ (blue) and $\Teq$ (green) for the whole $5\;a_0$ data series shown in Fig.~2(b) of the main paper. The solid black line shows the equilibrium $T_c$. }
\label{fig:T}
\end{figure}

First, for the thermal component we show in Fig.~S\ref{fig:T} that the radial velocity distribution still looks like a thermal distribution at a temperature very close to the calculated $\Teq(N,E)$. Here we show data for the same $5\;a_0$ series as shown in Fig.~2(b) in the main paper. In Fig.~S\ref{fig:T}(a) we show the distribution measured at $t=45\;$s, i.e. deep in the superheated regime. The data (red) is fitted almost perfectly by an equilibrium thermal distribution constrained to be characterised by $N, E$ and $\Teq$ (green). An unconstrained fit (blue) gives only a very slightly different shape with $T^f$ within few \% of $\Teq$. In Fig.~S\ref{fig:T}(b) we compare the calculated $\Teq$ (green) and the fitted $T^f$ (blue) for the whole $5\;a_0$ series. For comparison we also show the equilibrium $T_c$ (black line). Note that this is the same plot as in Fig.~2(b) in the main paper, with just the $T^f$ points added.

Second, for the collective excitations in the BEC to be in equilibrium with the thermal bath, the Landau-damping time $\tau_L$~\cite{Fedichev:1998SI, Pethick:2002SI} must be short compared to the characteristic time scale of the experiment.
For a uniform system at a temperature higher than the interaction energy per particle \cite{Pethick:2002SI}:

\begin{equation}
\frac{\tau_L \omega}{2\pi} \approx  \frac{n_0^{1/2}\lambda_T^2}{4 \pi a^{1/2}} \, ,
\label{eq:Landau}
\end{equation}
where $\omega$ is the excitation frequency, $n_0$ is the condensate density, and $\lambda_T$ is the thermal wavelength. For our assessment we use this uniform-system result with our peak $n_0$; this only overestimates $\tau_L$ for a harmonically trapped gas~\cite{Fedichev:1998SI}. 

Exactly at $a=0$ the damping time diverges and our theoretical picture breaks down. However, already for $a=1\;a_0$, for all our experimental parameters the RHS of Eq.~(\ref{eq:Landau}) is $<100$. Then, even for our lowest-energy modes, with $\omega/2\pi \sim 100\;$Hz, we get $\tau_L < 1\;$s.

\subsection{Non-equilibrium evolution of $N_0$}

The non-equilibrium evolution of $N_0$ is described by the differential equation
\begin{equation}
\dot{N_0}=-\kappa-\Gamma_0 N_0 \, ,
\label{eq:diff}
\end{equation}
where $\kappa$ is the coupling to the thermal bath due to elastic collisions and $\Gamma_0$ is the instantaneous inelastic loss rate per particle.
Following \cite{Gardiner:1997aSI} we use
\begin{equation}
\kappa=A \frac{8m(a \kB T)^2}{\pi \hbar^3}e^{2 \beta( \mu'-\mu_c)} \left [ e^{\beta(\mu_0-\mu')} -1 \right ] N_0 = A \gamma_{\rm el} N_0 \left [ e^{\beta (\mu_0-\mu_c)} - e^{\beta (\mu'-\mu_c)} \right ] \, ,
\label{eq:elasticterm}
\end{equation}
where $\gamma_{\rm el}=\frac{8m(a \kB T)^2}{\pi \hbar^3}e^{\beta (\mu'-\mu_c)}$ is essentially the elastic scattering rate for a thermal cloud at $\mu'$ and
$A\approx 1-10$ is a theoretically uncertain prefactor \cite{Gardiner:1998aSI}.

In the inelastic loss term we include contributions from one-body scattering and three-body recombination, $\Gamma_0=\Gamma_0^{(1)}+\Gamma_0^{(3)}$.
In our system, one-body loss is dominated by spontaneous scattering of photons from the trapping laser beams. We calculate it from the known wavelength and intensity  of the beams. For all the reported experiments
\begin{equation}
\Gamma_0^{(1)} \approx \frac{1}{35} \mathrm{s}^{-1} \, .
\label{eq:1body}
\end{equation}
The loss rate of condensate atoms due to three-body recombination in the presence of a thermal cloud is given by \cite{Soding:1999SI}
\begin{equation}
\Gamma_0^{(3)}=\frac{K_3(a)}{6} \left( \langle n_0^2 \rangle+6\langle n_0 n' \rangle+6 \langle n'^2 \rangle \right) \, ,
\label{eq:3body}
\end{equation}
where $n_0$ is the condensate density, $n'$ the thermal density, $K_3(a)$ the known $a$-dependant three-body coefficient \cite{Zaccanti:2009SI,Zaccanti:2007thSI}, and  $\langle... \rangle$ stands for an average over the density distribution. We set $n'$ to its value in the centre of the trap and for the condensate we again use a modified Thomas-Fermi approach:
\begin{equation}
\langle n_0\rangle =\frac{ \langle n_0\rangle_{\rm GS} }{ \left [1+ \left ( \langle n_0\rangle_{\rm GS}/\langle n_0\rangle_{\rm TF} \right )^{5/3} \right ]^{3/5}}  \, ,
\label{eq:n}
\end{equation}
\begin{equation}
\langle n_0^2\rangle =\frac{ \langle n_0^2 \rangle_{\rm GS} }{ \left [1+ \left ( \langle n_0^2 \rangle_{\rm GS}/\langle n_0^2 \rangle_{\rm TF} \right )^{5/6} \right ]^{6/5}}  \, .
\label{eq:n2}
\end{equation}
Here GS refers to the non-interacting Gaussian ground state and TF to the Thomas-Fermi approximation,
$ \langle n_0\rangle_{\rm TF} = \sqrt{2\pi} (15/7) (15 N_0 a / a_{\rm osc} )^{-3/5} \langle n_0\rangle_{\rm GS} $  and $\langle n_0^2\rangle_{\rm TF} =  \sqrt{3} \pi (15^2/56) (15 N_0 a / a_{\rm osc} )^{-6/5}  \langle n_0^2\rangle_{\rm GS}$, where
$ \langle n_0\rangle_{\rm GS} = N_0 / (2\pi a_{\rm osc}^2)^{3/2}$ and $\langle n_0^2\rangle_{\rm GS} = N_0^2 / (3 \pi^2 a_{\rm osc}^4)^{3/2}$.
Eqs.~(\ref{eq:n}) and (\ref{eq:n2}) smoothly interpolate between the ground state result (for $N_0a/a_{\rm osc} \ll 1$) and the Thomas-Fermi approximation (for $N_0a/a_{\rm osc} \gg 1$). Note that other forms which smoothly interpolate between these two limits give essentially the same results.

Using Eqs. (\ref{eq:diff}) - (\ref{eq:n2}) we simulate the evolution of the condensate atom number, $N_0(t)$, from its initial value $N_0(t=0)$.  We use the measured $N(t)$ and $E(t)$ and the numerically evolved $N_0(t)$ to obtain $\mu'(t)$ for use in Eq. (\ref{eq:elasticterm}).
%Having calculated $N_0(t)$ we obtain predictions for $\bar{t}$ and $\bar{N}$.
To determine $\bar{t}$ from our calculations we define the condensate to be present if $N_0$ is larger than $N_0^{\rm min}=3\kB T/(\hbar \omega)$, the thermal occupation of the first excited state.

\begin{figure} [h]
\includegraphics[width=0.5\columnwidth]{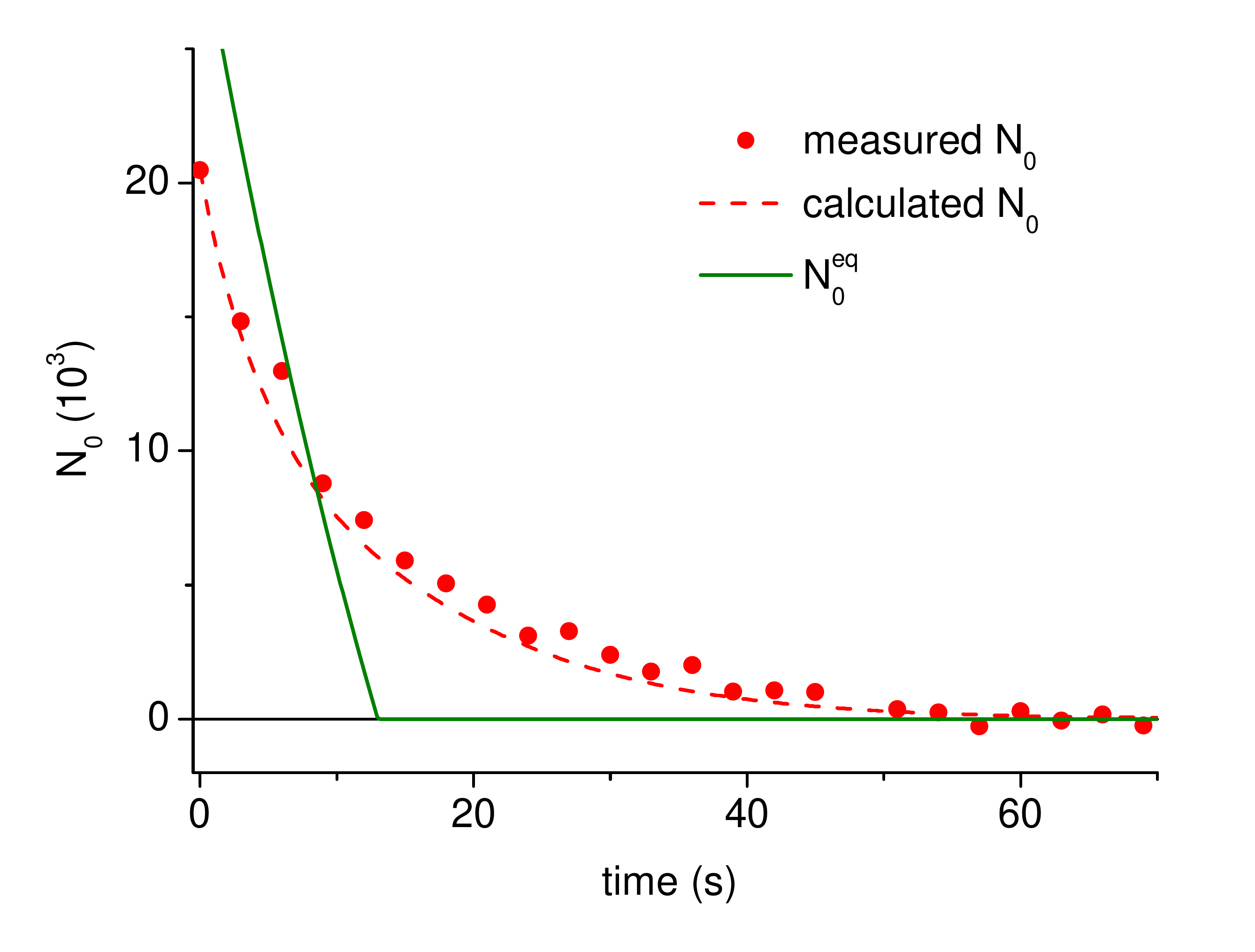}
\caption{\textbf{Non-equilibrium $N_0$ dynamics}. We plot the calculated $N_0(t)$ (dashed red line) together with the measured $N_0$ (red points) for the same $5\,a_0$ data series as in Fig.~2(b)  in the main text. For comparison we also show the calculated $\Neq(t)$ (solid green line).}
\label{fig:N0}
\end{figure}

In the main text we show the results of our calculations for $\bar{T}$ and $\bar{t}$ (Fig.~4), with $A=3$.
In Fig.~S\ref{fig:N0} we show that our calculations (with the same value of $A$) also describe well the full dynamics $N_0(t)$.

%\bibliography{Quench}

\begin{thebibliography}{1}
\bibitem{g2SI}
In Eqs.~(\ref{eq:N'}) and (\ref{eq:E}) we also replace the function ${\rm g}_2$ with a slightly modified dilogarithm function $\tilde{\rm g}(z)= (3/4) |{\rm g}_2 (z)|+ (1/4) {\rm Re} \left\{ {\rm g}_2 (z) \right\}$, which remains real for $z>1$. This is done just for mathematical convenience and does not affect our results since the integrals are strongly dominated by the region where $x^2/2 > (\mu'-\mu_c)/\kB T$ and $\tilde{\rm g} = {\rm g}_2$.
\end{thebibliography}
%\bibliographystyle{nature}

\end{document}